\documentclass[hyper,12pt,letterpaper]{JHEP3}

\usepackage{graphicx}
\usepackage{latexsym,amsmath,amsfonts,amssymb}
\usepackage{mathrsfs}
\usepackage[makeroom]{cancel}
\usepackage{bbm}
\usepackage{bm}
\usepackage{subfigure}
\usepackage{paralist}
\usepackage{url}

\numberwithin{equation}{section}

\newcommand{\be}{\begin{equation}} \newcommand{\ee}{\end{equation}}
\newcommand{\bea}{\begin{equation} \begin{aligned}} \newcommand{\eea}{\end{aligned} \end{equation}}

\newcommand{\pf}{\mathrm{Pf}}

\newcommand{\cH}{\mathcal{H}}

\newcommand{\cN}{\mathcal{N}}

\newcommand{\cQ}{\mathcal{Q}}
\newcommand{\cR}{\mathcal{R}}

\newcommand{\bC}{\mathbb{C}}

\newcommand{\bZ}{\mathbb{Z}}

\newcommand{\unit}{\mathbbm{1}}

\def\sl{\mathfrak{sl}}

\def\repa{\raise4pt\hbox{$\square$}\mkern-14mu\raise-4pt\hbox{$\square$}}
\def\repab{\overline{\raise4pt\hbox{$\square$}\mkern-14mu\raise-4pt\hbox{$\square$}\mkern-1mu}}

\DeclareMathOperator{\Tr}{Tr}

\newcommand{\PE}{\mathop{\rm PE}}

\title{The Hilbert series of $3d$ $\cN=2$ Yang-Mills theories with vectorlike matter}

\author{Stefano~Cremonesi\\

Department of Mathematics, King's College London \\
The Strand, London WC2R 2LS, United Kingdom
}

\abstract{This paper presents a formula for the Hilbert series that counts gauge invariant chiral operators in 3d $\cN=2$ Yang-Mills theories with vectorlike matter and no Chern-Simons interactions. The formula counts 't Hooft monopole operators dressed by gauge invariants of a residual gauge theory of massless fields in the monopole background, which is determined by the Higgs mechanism. The sum over magnetic charges is restricted due to instanton effects that partially lift the classical Coulomb branch. The formalism is applied to unitary and symplectic gauge theories with fundamental matter, reproducing old results for the moduli space of vacua and the chiral ring, without resorting to any further effective superpotential on the moduli space. 
}

\begin{document}

\section{Introduction}

The study of moduli spaces of supersymmetric vacua and the associated chiral rings has a long tradition of providing exact results on strongly coupled supersymmetric gauge theories (see \cite{Intriligator:1995au,Strassler:2003qg,Argurio:2003ym} for reviews), expecially in the context of four-dimensional field theories. 

Moduli spaces of vacua of supersymmetric gauge theories in three dimensions are subtler, because gauge fields can be dualized into periodic scalars which lead to new flat directions. While the dualization can be done explicitly for abelian gauge fields, its extension to non-abelian gauge fields is a long-standing open problem. Most results on moduli spaces of non-abelian supersymmetric gauge theories in three dimensions were thus obtained using semiclassical analysis in weakly coupled regions of the Coulomb branch of the moduli space where the gauge group is broken to its Cartan subgroup \cite{deBoer:1996mp,deBoer:1996ck,deBoer:1997kr,Aharony:1997bx,Aharony:1997gp,Karch:1997ux}. The dual photons for the Cartan subgroup are periodic and must be exponentiated to produce well-defined variables. Their insertion modifies the boundary conditions of fields in the path integral, introducing a magnetic flux around the insertion point. Local operators of this kind are called 't Hooft monopole operators \cite{'tHooft:1977hy}.

More recently, following the pioneering works \cite{Borokhov:2002ib,Borokhov:2002cg,Borokhov:2003yu} that constructed 't Hooft monopole operators well defined even at the origin of the moduli space, it has become clear that the analysis of monopole operators offers a powerful alternative to the dualization of non-abelian gauge fields, and allows to obtain exact results not only on global symmetries \cite{Gaiotto:2008ak,Bashkirov:2010kz,Bashkirov:2010hj} but more generally on the moduli space of vacua and chiral ring of 3d $\cN=4$ supersymmetric gauge theories \cite{Cremonesi:2013lqa,Cremonesi:2014kwa,Cremonesi:2014vla,Cremonesi:2014xha,Cremonesi:2014uva, Bullimore:2015lsa}.
The key ingredient in the new understanding of the moduli space of vacua of 3d $\cN=4$ gauge theories has been an exact formula --- referred to as the \emph{monopole formula} --- for the Hilbert series of the Coulomb branch of their moduli space \cite{Cremonesi:2013lqa}.%
\footnote{The monopole formula of \cite{Cremonesi:2013lqa} has recently sparked interesting progress towards a mathematical definition of the Coulomb branch of 3d $\cN=4$ gauge theories  \cite{Nakajima:2015txa}.}  The Hilbert series is a generating function that counts bosonic gauge invariant chiral operators, which are annihilated by two supercharges $\bar{\cQ}_\alpha$ of a 4-supercharges superalgebra.
The monopole formula for the Hilbert series of the Coulomb branch of a 3d $\cN=4$ theory counts $\cN=2$ chiral monopole operators dressed by the adjoint chiral that arises in the decomposition of the $\cN=4$ vector multiplet into $\cN=2$ multiplets. By standard techniques it is possible to extract from the Hilbert series information on the quantum numbers of the generators and of the relations of the chiral ring. The chiral ring relations involving monopole operators are purely quantum relations and do not follow from a superpotential. 

The purpose of this paper is to extend the monopole formula to moduli spaces of 3d $\cN=2$ gauge theories without bare or effective Chern-Simons interactions. Matter fields are therefore taken to transform in vectorlike representations and to have vanishing bare real masses.%
\footnote{By vectorlike matter I mean a matter content which does not lead to chiral gauge anomalies in four dimensions, and in three dimensions does not lead to Chern-Simons interactions in the effective description on the Coulomb branch, where matter fields gain a mass and are integrated out.} All monopole operators are thus neutral under the gauge group. 
The monopole formula for the Hilbert series of the moduli space of such 3d $\cN=2$ gauge theories also counts dressed chiral monopole operators. In this case the dressing is due to the matter fields of the theory. More precisely, the background for a bare gauge variant BPS monopole operator of magnetic charge $m$ leaves a residual gauge group $H_m$ of massless vector multiplets and a residual matter content of massless chiral multiplets. This defines a \emph{residual gauge theory} of the massless fields. Gauge invariant dressed monopole operators are constructed by dressing the bare gauge variant monopole operators with gauge invariants of the residual gauge theory, and then averaging over the Weyl group of $G$ to obtain fully gauge invariant objects. 

This discussion at the level of monopole operators defined by singular boundary conditions in the path integral has a direct counterpart in the old description of monopole operators on the classical moduli space. When the real scalar $\sigma$ in the 3d $\cN=2$ vector multiplet acquires a VEV, exploring the real Coulomb branch base of the classical moduli space, it gives mass to some vector multiplets and chiral multiplets as in the Higgs mechanism. This leaves a residual gauge theory of massless fields, which may have a Higgs branch parametrized by gauge invariant operators such as mesons and baryons. The Higgs branch of the residual gauge theory is foliated over the real Coulomb branch. Finally, the periodic dual photons $\tau$ complexify $\sigma$ into $\Phi=\sigma + i \tau$, which has to be exponentiated to give a single-valued  monopole operator. $\tau$ is a fiber coordinate over the space discussed before. 1-loop corrections control how the dual photon $\tau$ is fibered. See \cite{Aganagic:2009zk,Benini:2009qs,Benini:2011cma} for a discussion of this fibration in the context of 3d $\cN=2$ theories on the worldvolume of M2-branes and a review of how the Higgs branch leaves over $\sigma$-space can be obtained as a symplectic quotient.

The advantage of monopole operators defined via singular boundary conditions is that they account for quantum corrections and provide a description that is valid even in strongly coupled regions of the moduli space with unbroken non-abelian gauge symmetry, where the old semiclassical analysis does not apply.

In view of the previous discussion, one would naively count dressed monopole operators labelled by all magnetic charges in a Weyl chamber of the GNO lattice. This, however, ignores non-perturbative corrections to the superpotential on the Coulomb branch, which are generated by three-dimensional instantons that descend from smooth monopole configurations in four dimensions \cite{Affleck:1982as}.  The dynamically generated superpotential lifts most of the Coulomb branch, typically leaving a one-dimensional quantum Coulomb branch for a simple non-abelian gauge group. I will argue in the next section that the non-perturbative lifting of the Coulomb branch is realized in the monopole formula by restricting the summation over magnetic charges to a sublattice of the fundamental Weyl chamber of the GNO lattice, that is defined by the same inequalities that define the unlifted Coulomb branch in terms of the real scalar $\sigma$. 

The plan of the paper is as follows. In section \ref{sec:_Hilbert_series}, I review the concept of Hilbert series, discuss BPS monopole operators, dressing and lifting, and finally present the monopole formula \eqref{monopole_formula} for the Hilbert series of the moduli space of a 3d $\cN=2$ Yang-Mills theory with vectorlike matter. In sections \ref{sec:U(1)_flavors}--\ref{sec:USp(2N)_Nf}, I evaluate the monopole formula for the Hilbert series and deduce generators and relations of the chiral ring for $U(1)$, $SU(2)$, $U(N)$ and $USp(2N)$ gauge theories with fundamental matter. The results reproduce the picture of the moduli space of supersymmetric vacua obtained in \cite{deBoer:1997kr,Aharony:1997bx,Karch:1997ux,Aharony:1997gp}, without using any (regular or singular) effective superpotential. The quantum relations involving monopole operators are simply a consequence of the Coulomb-Higgs fibration structure of the moduli space. I close the paper with some conclusions in section \ref{sec:_conclusions}.

\vskip 0.3cm
\noindent
{\it Note added:}  During the completion of this paper I became aware of a related work \cite{competitors} that studies moduli spaces of 3d $\cN=2$ theories using similar methods. I thank the authors for sharing with me a draft prior to publication.

\section{The Hilbert series of the moduli space of 3d $\cN=2$ Yang-Mills theories with vectorlike matter}
\label{sec:_Hilbert_series}

The Hilbert series is a generating function
\be\label{Hilbert_series_def}
H(t,\mathbf{x})= \Tr_\cH( t^R \prod_I \mathbf{x}_I^{F_I})
\ee
that counts bosonic gauge invariant chiral operators, which are annihilated by two  supercharges $\bar{\cQ}_\alpha$. $\cH$ denotes the Hilbert space of such chiral operators, which are graded according to their quantum numbers, namely an $R$-charge and other global charges $F_I$ that commute with the supercharges.
Since gauge invariant chiral operators correspond to holomorphic functions on the moduli space of vacua, the Hilbert series provides a quantitative characterization of the geometry of the moduli space. 

An alternative quantity that counts protected operators is the R\"omelsberger (or ``superconformal'') index \cite{Romelsberger:2005eg}, which is the supersymmetric partition function on $S^1\times S^{d-1}$ \cite{Festuccia:2011ws}. Unlike the Hilbert series, the R\"omelsberger index is a supertrace, it also counts non-chiral operators and is only sensitive to the superpotential via the $R$-charges of matter fields. The focus of this paper is on the moduli space of supersymmetric vacua and the chiral ring, therefore the Hilbert series is the natural generating function to look at. 
Note that since the Hilbert series of the full moduli space and the R\"omelsberger index depend on the same number of fugacities, the former cannot be obtained as a limit of the latter (that is however possible for sub-branches of the moduli space \cite{Razamat:2014pta}).

Hilbert series have been computed for a variety of 4d $\cN=1$ supersymmetric gauge theories \cite{Benvenuti:2006qr,Feng:2007ur,Gray:2008yu,Hanany:2008kn} (see also \cite{Pouliot:1998yv} for an early appearance). Assuming for simplicity of presentation that there is no superpotential, the Hilbert series of these theories is given by the Molien-Weyl formula
\be\label{Molien_def}
H(t,\mathbf{x}) = \int d\mu_{G}(w) \PE[\sum_a t^{R_a} \chi_{\cR^G_a}(w) \chi_{\cR^F_a}(\mathbf{x})]~,
\ee
where the sum inside the $\PE$ symbol runs over all chiral multiplets in the theory, and the integral is over the maximal torus of the gauge group $G$, with the Haar measure $d\mu_{G}$. 
$R_a$ is the $R$-charge of the $a$-th chiral multiplet $X_a$, which transforms in the representation $\cR^G_a$ of the gauge group $G$ and $\cR^F_a$ of the global (non-R) symmetry group $G_F$, and $\chi_\cR$ denotes the character of the representation $\cR$.
The integrand is the generating function of gauge variant chiral operators: in the absence of a superpotential, it simply counts all monomials in the chiral operators.%
\footnote{In the presence of a superpotential $W(X)$, the integrand is given by the $F$-flat Hilbert series, that is the Hilbert series of the ring of gauge variant chiral operators modulo the ideal of $F$-term equations, $\bC[X]/\langle dW(X) \rangle$. When the $F$-flat moduli space is a complete intersection, the $F$-flat Hilbert series has a simple plethystic exponential expression. When it is not a complete intersection, its Hilbert series can be computed using algebraic geometric software such as {\tt Macaulay2} \cite{M2}.}
The plethystic exponential $\PE$ of a multivariate function $f(t_1, . . . , t_n)$ such that $f(0,...,0) = 0$ is the generating function of symmetrizations, defined as 
\be\label{PE_def}
\PE \left[ f(t_1, t_2, \ldots, t_n) \right] = \exp \left( \sum_{k=1}^\infty \frac{1}{k} f(t_1^k, \cdots, t_n^k) \right)~.
\ee
In particular $\PE[\sum_a n_a \prod_i t_i^{m_a}]=\prod_a (1-\prod_i t_i^{m_a})^{-n_a}$, and expanding the geometric series shows that the integrand of \eqref{Molien_def} counts all monomials in the chiral multiplets $X_a$, which are bosonic variables and hence symmetrized. The integral with the Haar measure finally projects to gauge invariant combinations.

Much like the Hilbert series of the Coulomb branch of 3d $\cN=4$ theories \cite{Cremonesi:2013lqa}, the Hilbert series of the moduli space of 3d $\cN=2$ thories takes a different form from the Molien formula \eqref{Molien_def} for the moduli space of 4d $\cN=1$ theories. The novelty is that the gauge invariant chiral operators include 't Hooft monopole operators, that acquire expectation value on the Coulomb branch.

\subsection{Bare monopole operators}\label{subsec:bare_monopole}

A monopole operator is not a polynomial in the microscopic fields that are path integrated over in the quantum theory. It is instead a local disorder operator, that can be defined by prescribing a boundary condition for the path integral at its insertion point. A \emph{bare} monopole operator $V_m(x)$ is defined by requiring that the gauge fields have a Dirac monopole singularity (specified by an embedding $U(1)\hookrightarrow G$) at the insertion point $x$,
\be\label{DiracMonopole}
A_\pm \sim \frac{m}{2}(\pm 1 - \cos\theta) d\varphi~,
\ee
where $(r,\theta,\varphi)$ are spherical coordinates around $x$ and $A_\pm$ is the gauge connection in the northern/southern patch of the two-sphere surrounding $x$. The magnetic charge $m$ belongs to the Lie algebra $\mathfrak{g}$ of the gauge group $G$. It can be gauge rotated in each patch to a constant element of the Cartan subalgebra $\mathfrak{t}\subset \mathfrak{g}$, defined modulo the action of the Weyl group $W_G$.
Dirac quantization \cite{Englert:1976ng} 
\be\label{Dirac_quant}
\exp (2\pi i m) = \unit_G\;
\ee
requires that the magnetic charge $m$ belong to the weight lattice $\Gamma_{G^\vee}$ of $G^\vee$, the GNO (or Langlands) dual group of the gauge group $G$ \cite{Goddard:1976qe}. 
Therefore monopole operators for the gauge group $G$ are labelled by magnetic charges $m$ which are weights of the dual group $G^\vee$, and gauge invariant monopole operators by magnetic charges $m$ taking values in the quotient space $\Gamma_{G^\vee}/W_{G}$ \cite{Kapustin:2005py}. 

The bare monopole operator $V_m$ defined by the boundary condition \eqref{DiracMonopole} becomes the lowest component of an $\cN=2$ chiral multiplet if the boundary condition
\be\label{sing_sigma}
\sigma \sim \frac{m}{2r}
\ee
for the non-compact real scalar $\sigma$ in the $\cN=2$ vector multiplet is also imposed.  \eqref{DiracMonopole}, \eqref{sing_sigma} satisfy the BPS equation $ (d-iA) \sigma = -\ast F$ and preserve the same supersymmetries of an $\cN=2$ chiral multiplet \cite{Borokhov:2003yu}. It was shown in \cite{Borokhov:2002cg} tht there is a unique BPS bare monopole operator per magnetic charge $m$ in $\Gamma_{G^\vee}/W_{G}$. Note that the boundary conditions that define the gauge variant monopole operator of magnetic charge $m$ break the gauge group $G$ to a residual gauge group $H_m$, the commutant of $m$ in $G$, by the adjoint Higgs mechanism. The roots of the Lie algebra $\mathfrak{h}_m$ of $H_m$ are those roots $\alpha$ of the Lie algebra $\mathfrak{g}$ of $G$ such that $\alpha(m)=0$.

Bare monopole operators can be charged under the topological symmetry group $G_J=Z(G^\vee)$, the center of the GNO dual group \cite{Kapustin:2005py}. I will denote by $J(m)$ the topological charge of a monopole operator of magnetic charge $m$, and by $z$ the associated $G_J$-valued fugacity. For instance, if the gauge group is $G=U(N)$ with magnetic charge $\mathrm{diag}(m_1,\dots,m_N)$, the topological symmetry is $G_J=U(1)$ and $J(m)=\sum_i m_i$. 

Monopole operators also acquire charges under the other global ($R$- or flavor) symmetries of the theory at the quantum level. If $Q$ is the global charge in question, the charge $Q(m)\equiv Q[V_m]$ of the bare monopole operator $V_m$ is \cite{Borokhov:2002cg, Imamura:2011su,Benini:2011cma}
\be\label{monopole_charge}
Q(m)=-\frac{1}{2}~\sum_{\mathrm{fermi}~\psi_i}~ \sum_{\rho_i \in \cR^G_i} ~Q[\psi_i] \,|\rho_i(m)| ~,
\ee
where the sum is over all fermions $\psi_i$ in the theory, $Q[\psi_i]$ is the global charge of the fermion $\psi_i$ and $\rho_i$ is the weight of $\psi_i$ under the gauge group $G$. 

Taking into account all charges, a bare monopole operator $V_m$ is counted in the Hilbert series with the weight 
\be\label{weight_bare_monopole}
t^{R(m)} z^{J(m)} \prod_i x_i^{F_i(m)}~,
\ee
where $(\mathbf{x}_I)=(z,x_i)$ and $(F_I)=(J,F_i)$ in the notation of \eqref{Hilbert_series_def}.

\subsection{Dressed monopole operators}\label{subsec:dressed_monopole}

Let us now turn to the dressing of monopole operators. In the definition of the bare monopole operators considered so far, all matter fields vanish. It is easy to see that constant matter fields can be turned on without affecting the supersymmetry of the monopole operator, as long as $\rho(m)=0$ for the weight $\rho$ of the representation of the matter field under the gauge group.%
\footnote{In the presence of a superpotential, the $F$-term equations also need to be satisfied.}
I will refer to such matter fields as ``massless'' with an abuse of terminology, by analogy with the moduli space description where $\sigma$ acquires a constant VEV and gives an effective real mass $\rho(\sigma)$ to matter fields with weight $\rho$ under the gauge group. The relation between $\sigma$ and $m$ in the two pictures is due to the BPS boundary condition \eqref{sing_sigma}. 

In conclusion, gauge variant bare chiral monopole operator can be dressed by massless matter fields, which in turn transform under the residual gauge group $H_m$. In the following I will refer to the gauge theory with massless vector multiplets for the gauge group $H_m$ and matter content given by the massless matter fields in the monopole background as the \emph{residual gauge theory} $T_m$ associated to the bare monopole operator $V_m$ or equivalently to the magnetic charge $m$. Due to the adjoint Higgs mechanism, the residual gauge group $H_m$ has the same rank as $G$ and often includes factors with no charged matter in $T_m$. In the rest of the paper I will often omit these decoupled factors in $T_m$, which play no role in the analysis since their only gauge invariant is the identity operator.

Finally, to impose gauge invariance under $G$, the gauge variant bare monopole operator  dressed by $H_m$-gauge invariant chiral operators of the residual gauge theory $T_m$ is averaged over the Weyl group of $G$. 

The dressing by gauge invariants of the residual gauge theory $T_m$ is implemented by multiplying the weight \eqref{weight_bare_monopole} for the bare monopole operator
by a dressing factor that counts gauge invariants of $T_m$. Therefore this dressing factor is nothing but the Hilbert series $H_{T_m}(t,x)$ of the residual gauge theory $T_m$, that is given by a Molien-Weyl formula like \eqref{Molien_def}:
\be\label{dressing}
H_{T_m}(t,x_i) = \int d\mu_{H_m}(w) \PE[\sum_{a| X_a \in T_m} t^{R_a} \chi_{\cR^{H_m}_a}(w) \chi_{\cR^F_a}(x)]~,
\ee
where the integrand only includes the massless matter content of the residual gauge theory $T_m$, \emph{i.e.} matter fields with $\rho(m)=0$, and the integral is over the residual gauge group $H_m$.     
In writing \eqref{dressing} I assumed again for technical simplicity that the theory has no superpotential, although this assumption can be relaxed.

The dressing factor $H_{T_m}$ can also be written generally as an integral over $G$ as in \eqref{Molien_def}, if each multiplicative factor due to a matter field of weight $\rho$ in the integrand is raised to the power $\delta_{\rho(m),0}$, and similarly each factor due to a root $\alpha$ in the Haar measure is raised to the power $\delta_{\alpha(m),0}$.

\subsection{Lifting by instanton generated superpotentials} \label{subsec:lifting}

As was anticipated in the introduction, this cannot be whole story. It is well known that BPS instantons in three dimensions, which are smooth BPS monopole configurations with magnetic charge in the coroot lattice of the gauge Lie algebra \cite{Weinberg:1979zt,Weinberg:1982ev}, 
can generate a superpotential  \cite{Affleck:1982as} which lifts most of the classical Coulomb branch quantum-mechanically \cite{deBoer:1997kr,Aharony:1997bx}. This must be taken into account when counting dressed monopole operators in the Hilbert series.

Due to fermionic zero modes in the instanton background, that can be counted using the Callias index \cite{Callias:1977kg},%
\footnote{See the appendix of \cite{deBoer:1997kr} for a nice account based on \cite{Weinberg:1979zt,Weinberg:1982ev}.}  the form of the dynamically generated superpotential depends on the specific sub-wedge of the Coulomb branch to which the real scalar $\sigma$ in the vector multiplet belongs \cite{deBoer:1997kr,Aharony:1997bx}. Let us recall how the subdivision into sub-wedges arises. By gauge symmetry, $\sigma$ can be taken to be in the fundamental Weyl chamber defined by $\alpha_a(\sigma)\geq 0$ for all simple roots $\alpha_a$. The instanton factors $Y_a^{-1} \simeq e^{-\alpha_a^\vee \cdot \Phi/g^2}$ for smooth BPS monopole configurations with magnetic charge given by a simple coroot have two gaugino zero modes regardless of the value of $\sigma$. Here  $\Phi=\sigma + i \tau$ and $g$ is the Yang-Mills coupling. The $\simeq$ symbol indicates that the relation is obtained using the classical action evaluated in the BPS instanton background (1-loop corrections will be discussed shortly). 
Therefore $Y_a^{-1}$ contributes a term to the dynamically generated superpotential if there are no further quark zero modes. 
The number of quark zero modes is  \cite{deBoer:1997kr}
\be\label{quark_zero_modes}
N_{quark}(n,\sigma)= \frac{1}{2} \sum_i \sum_{\rho_i \in \cR_i^G} \rho_i(n)
~\mathrm{sign}(\rho_i(\sigma))~,
\ee
where the sum is over all weights $\rho_i$ of matter field representations under the gauge group, and $n=\sum_{a=1}^{\mathrm{rk}{G}} n_a \alpha_a^\vee$ is the magnetic charge of the instanton ($n_a\geq 0$ for all $a$ for BPS instantons). Strictly speaking \eqref{quark_zero_modes} was derived when $\rho_i(\sigma)\neq 0$, but it can be extended to the case $\rho_i(\sigma)\neq 0$ with the definition $\mathrm{sign}(0)=0$ (this agrees with \cite{Aharony:1997bx}). 
Note that \eqref{quark_zero_modes} depends on $\sigma$ through the sign of the effective real mass $\rho_i(\sigma)$ of the matter fields.  It is therefore useful to subdivide the fundamental Weyl chamber in $\sigma$-space into sub-wedges according to the signs of the effective real masses of the matter fields: the number of fermionic zero modes of instanton configurations is constant in the interior of each sub-wedge. 
These sub-wedges in $\sigma$-space are in one-to-one correspondence with the domains of linearity of the charge formula \eqref{monopole_charge} for monopole operators in $m$-space.

In addition, it is important to include quantum corrections in the relation between $Y_a^{-1}$ and $\Phi$. One-loop corrections due to $W$-bosons make $Y_a^{-1}$ also proportional to a positive power of $\alpha_a(\sigma)$ \cite{Dorey:1998kq}, therefore no superpotential term is generated if $\alpha_a(\sigma)=0$.

For instance \cite{deBoer:1997kr,Aharony:1997bx}, for a $U(N)$ gauge theory with $N_f$ flavors of fundamental and antifundamental matter, the fundamental Weyl chamber is the closed cone $\sigma_1\geq \sigma_2 \geq \dots \geq \sigma_N$. This is subdivided into $N+1$ maximal dimensional sub-wedges according to the signs of $\sigma_a$. The number of quark zero modes \eqref{quark_zero_modes} of the $a$-th fundamental instanton, with $n=\alpha_a^\vee=e_a-e_{a+1}$, is $N_f(\mathrm{sign}(\sigma_a)-\mathrm{sign}(\sigma_{a+1}))$. Therefore $Y_a^{-1}$ contributes a term to the effective superpotential which lifts the classical Coulomb branch if and only if $\sigma_a$ and $\sigma_{a+1}$ have the same sign. The unlifted Coulomb branch is the complexification by the dual photon of $\sigma_1\geq \sigma_2 = \dots = \sigma_{N-1}=0 \geq \sigma_N$, which is two-dimensional. (One dimension comes from the $U(1)$ part of the gauge group, which has no instantons, the other comes from the $SU(N)$ part, that is partially lifted by instantons.) 

This semiclassical analysis, which is valid far out on the classical Coulomb branch, has a counterpart in the picture of monopole operators defined by boundary conditions in the path integral. By the correspondence between moduli space and chiral ring, the monopole operators that parametrize would-be flat directions that are lifted by the dynamically generated superpotential cannot be \emph{bona fide} chiral operators. They must be either non-chiral or possibly chirally exact, therefore vanishing in the chiral ring of the quantum theory, once non-perturbative corrections are included.

Since the BPS boundary condition \eqref{sing_sigma} relates the magnetic charge $m$ of the monopole operator to the scalar $\sigma$ in the vector multiplet, and the latter is forced to belong to the sub-cone of the classical Coulomb branch which survives non-perturbative corrections, it is natural to propose that the magnetic charge $m$ of \emph{bona fide} chiral monopole operators is further restricted by the same inequalities that restrict $\sigma$ on the Coulomb branch. In other words, I will restrict to magnetic charges belonging to the \emph{quantum sublattice}   
\be\label{quantum_sublattice_m}
\Gamma_q = \{m \in \Gamma_{G^\vee}/W_G~ |~ N_{quark}(\alpha_a^\vee, m) \neq 0 \vee \alpha_a (m)=0 ~\forall a=1,\dots,\mathrm{rk}(G) \}
\ee
of the fundamental Weyl chamber of the GNO lattice.

Alternatively, treating $\Phi=\sigma+i\tau$ as a chiral multiplet, one would impose its $F$-term equations from the instanton generated superpotential in order to satisfy the BPS equations preserved by the boundary condition, which together with \eqref{sing_sigma} leads to the same conclusion.

While this argument seems plausible and leads to results consistent with the analysis of moduli spaces using effective superpotentials valid away from the origin \cite{Aharony:1997bx,deBoer:1996ck}, it would be nice to make this point more precise. It would be interesting to see whether the method of equivariant integration introduced in \cite{Bullimore:2015lsa} in the context of 3d $\cN=4$ theories could be used to show that monopole operators with magnetic charges outside \eqref{quantum_sublattice_m} are non-chiral or trivial in the chiral ring.

\subsection{The monopole formula for the Hilbert series} \label{subsec:monopole_formula}

The previous discussion of bare monopole operators, dressed monopole operators and non-perturbative lifting of part of the Coulomb branch can be neatly summarized in a \emph{monopole formula} for the Hilbert series that counts gauge invariant dressed monopole operators that parametrize the moduli space of the 3d $\cN=2$ gauge theory:
\be\label{monopole_formula}
H(t,z,x_i) = \sum_{m \in \Gamma_q} \bigg[ t^{R(m)} z^{J(m)} \prod_i x_i^{F_i(m)} \bigg] \cdot H_{T_m}(t,x_i) ~.
\ee
The summation is over the sublattice $\Gamma_q$ \eqref{quantum_sublattice_m} labelling bona fide chiral monopole operators that parametrize the unlifted quantum Coulomb branch. The  factor in brackets is the weight \eqref{weight_bare_monopole} associated to a bare monopole operator. The final factor is the dressing factor \eqref{dressing} by gauge invariants of the residual gauge theory. Formula \eqref{monopole_formula} is the main result of this paper.

In order for the Hilbert series \eqref{monopole_formula} to be a Taylor series in $t$, I require that there exists a $U(1)_R$ symmetry which gives positive charge to all gauge invariant chiral operators.
In the following I will restrict to gauge theories that satisfy this condition, so that there are finitely many chiral gauge invariant operators at any fixed value of the $R$-charge. In particular the number of flavors will be above the bound that gives a quantum deformation of the moduli space \cite{Aharony:1997bx,Karch:1997ux}.

In the case of superconformal field theories, the superconformal $R$-symmetry (which can be computed by $F$-maximization \cite{Jafferis:2010un,Closset:2012vg} in the absence of accidental symmetries) must assign charge greater than or equal to $1/2$ to  gauge invariant chiral operators by the unitarity bound. Since this paper is concerned with the moduli space and chiral ring of supersymmetric gauge theories, I will not worry about the superconformal R-symmetry, nor accidental symmetries and decoupling of free fields that are expected to arise if no UV $R$-symmetry satisfies the stricter unitarity bound. In particular I will be able to study SQCD theories with $U(N)$ gauge group and $N_f=N$ fundamental flavors, which violate the unitarity bound for $N>2$ but admit an $R$-symmetry with positive charges.

In the rest of the paper I will evaluate the monopole formula \eqref{monopole_formula} for various families of 3d $\cN=2$ gauge theories with vectorlike matter and no Chern-Simons interactions. For simplicity I will restrict to theories without baryons nor bare superpotential interactions (both assumptions can be relaxed). 
The results reproduce the old analysis of the moduli space based on semiclassical analysis and an effective superpotential which involve monopole operators and mesonic operators \cite{Aharony:1997bx,Karch:1997ux,Aharony:1997gp}. This effective superpotential is typically singular at the origin of the moduli space, and requires to constrain the maximal rank of mesons based on semiclassical analysis to deduce the moduli space. In the formalism of this paper, the results are obtained without invoking any effective superpotential. This is important because there is no guarantee that the quantum relations in the chiral ring should follow from a superpotential: for instance there is no superpotential that implies the quantum relations on the Coulomb branch of 3d $\cN=4$ theories. In addition the generators of the chiral ring could involve dressed monopole operators which cannot be expressed in terms of gauge invariant bare monopole operators and gauge invariants that only involve matter fields, even though this does not happen for the simple theories considered in the rest of the paper. The current formalism can also account for such situations, as was seen  in \cite{Cremonesi:2013lqa} in the context of 3d $\cN=4$ theories.

The Hilbert series \eqref{monopole_formula} is computed by simply counting dressed monopole operators.%
\footnote{It is crucial in this counting that there exists 
a unique BPS bare monopole operator per magnetic charge \cite{Borokhov:2003yu}.} 
Information on the quantum numbers of generators and relations in the chiral ring can be extracted from the Hilbert series using plethystic techniques \cite{Feng:2007ur}. In simple examples like the theories that will be studied in the following sections, the Coulomb-Higgs branch fibration structure of the moduli space and of the Hilbert series is in fact enough to identify generators and relations.

\section{$U(1)$ with $N_f\geq 1$ flavors}\label{sec:U(1)_flavors}

Consider first $3d$ $\cN=2$ SQED with $N_f\geq 1$ flavors of massless electrons $Q$ and $\tilde{Q}$ of charge $+1$ and $-1$. The charges of $Q$, $\tilde{Q}$, the meson $M$ and the fundamental monopole operators $V_+\equiv V_{1}$ and $V_-\equiv V_{-1}$ are listed in table \ref{tab:charges_U(1)}.   
\begin{table}[h]
\centering
\begin{tabular}{c| c c c c}
         &  $U(1)_R$ & $U(1)_A$  &  $SU(N_f)_L \times SU(N_f)_R$  & $U(1)_J$ \\ \hline
$Q$    &   $r$  &  $1$   & $[0,0,\dots,0,1; 0,0,\dots,0,0]$ &   $0$  \\
$\tilde{Q}$  & $r$  &  $1$  & $[0,0,\dots,0,0;1,0,\dots,0,0]$  &   $0$  \\ \hline
$M$    & $2r$  &  $2$   & $[0,0,\dots,0,1;1,0,\dots,0,0]$ &   $0$  \\
$V_+$  & $N_f(1-r)$  &  $-N_f$   & $[0,0,\dots,0,0;0,0,\dots,0,0]$ &  $1$ \\
$V_-$  & $N_f(1-r)$  &  $-N_f$   & $[0,0,\dots,0,0;0,0,\dots,0,0]$ &  $-1$ \\
\hline
fugacities & $t$ & $y$  & $u\qquad~,~\qquad v$ & $z$
\end{tabular}
\caption{Charges under the global symmetry for $U(1)$ with $N_f$ flavors. Representations of non-abelian groups are denoted by their Dynkin labels.}
\label{tab:charges_U(1)}
\end{table}

The nontrivial charges of a general monopole operator are 
\be\label{charges_monopole_U(1)}
R[V_{m}] =N_f(1-r)|m| ~, \qquad
A[V_{m}] =-N_f ~, \qquad  J[V_{m}] = m~.
\ee

The $R$-charge $r$ of the matter fields is taken to satisfy $0<r<1$, so that all gauge invariant operators have positive $R$-charge (if $N_f>0$) and the Hilbert series is a Taylor series in $t$. (For $N_f=0$ monopole operators have zero $R$-charge and the moduli space is the classical Coulomb branch, which is a cylinder.)

Since the gauge group is abelian, there are no instantons. The moduli space of vacua of the theory is known to split into three components generated respectively by $V_+$ (the Coulomb component $m\geq 0$), $V_-$ (the Coulomb component $m\leq 0$) and $M$ (the mesonic component over $m=0$), which meet at the origin \cite{deBoer:1997kr,Aharony:1997bx}. 

This well-known result is reproduced by the computation of the Hilbert series using the monopole formula \eqref{monopole_formula}. For $m\neq 0$, all matter fields are massive and the sum is simply over bare monopole operators 
\be
V_{m} = \begin{cases} 
(V_+)^m & m>0 \\
(V_-)^{-m} & m<0 
\end{cases}~,
\ee
which contribute the weight $(t^{N_f(1-r)} y^{-N_f})^{|m|} z^m$ to the Hilbert series.

For $m=0$, the chiral gauge invariants are powers of the mesons $M=\tilde{Q} Q$, which have rank $1$ due to the $U(1)$ gauge contraction. The contribution to the Hilbert series is the mesonic Hilbert series of $U(1)$ with $N_f$ flavors,
\be\label{U1_mesonic}
H_{T_0}(t,y,u,v)= \sum_{n=0}^\infty [0,\dots,0,n;n,0,\dots,0]_{u,v}(t^r y)^{2n}~,
\ee
where the character $\chi_{[0,\dots,0,n;n,0,\dots,0]}(u,v)$ has been denoted as $[0,\dots,0,n;n,0,\dots,0]_{u,v}$ with an abuse of notation to improve readability.
Unrefined with respect to the flavor symmetry, the Hilbert series of the mesonic branch is 
\be\label{U1_mesonic_ur}
H_{T_0}(t,y,1,1,z)= {}_{2}F_1(N_f,N_f;1;t^{2r} y^2)~.
\ee

Adding the contributions from $m\geq 0$, $m\leq 0$ and $m=0$ with mesonic dressing, and subtracting $2$ not to overcount the identity operator corresponding to the origin, the Hilbert series of the total moduli space is
\be\label{U(1)_HS}
\begin{split}
H(t,y,u,v,z)&= \frac{1}{1-(t^{1-r} y^{-1})^{N_f} z} +\frac{1}{1-(t^{1-r} y^{-1})^{N_f} z^{-1}} + \\ 
&+\sum_{n=0}^\infty [0,\dots,0,n;n,0,\dots,0]_{u,v}(t^r y)^{2n} -2 ~.
\end{split}
\ee

The expression \eqref{U(1)_HS} exhibits the three components of the moduli space meeting at the origin, each of which contributes a term to the Hilbert series. It shows that the chiral ring is generated by the monopole operators $V_+$ and $V_-$ and by the $N_f\times N_f$ meson matrix $M$, subject to the relations
\be\label{U(1)_relations}
V_+ V_- = 0~, \quad V_+ M = 0~, \quad V_- M = 0~, \quad \mathrm{rk}(M)=1~.
\ee
The first three relations arise because the three components of the moduli space only meet at a point, whereas the fourth set of relations (shorthand for the vanishing of all $2\times 2$ minors of $M$) are the standard relations on the mesonic branch of the $U(1)$ gauge theory with $N_f$ flavors. This statement can be checked explicitly using {\tt Macaulay2}: the Hilbert series of the graded ring generated by $V_+$, $V_-$ and $M$ subject to \eqref{U(1)_relations} reproduces the formula \eqref{U(1)_HS} obtained by counting dressed monopole operators. 

It should be emphasized that the quantum relations involving monopole operators are an output of our analysis. They are not derived from an effective superpotential $W_{eff} = (V_+ V_- \det M)^{1/N_f}$ (which is singular for $N_f>1$) supplemented with the constraint $\mathrm{rk}(M)\leq 1$ as in \cite{Aharony:1997bx}.

For $N_f=1$ we recover the moduli space of the $XYZ$ model, the Wess-Zumino model with superpotential $W=V_+ V_- M$, which consists of three copies of $\bC$ glued at the origin. Choosing $r=1/3$ gives equal $R$-charge $2/3$ to the three generators.

\section{$SU(2)$ with $N_f\ge 2$ flavors}\label{sec:SU(2)_flavors}

Next we consider $3d$ $\cN=2$ SQCD with gauge group $SU(2)$ and $N_f\geq 2$ flavors. Since the doublet is a real representation, the $N_f$ quark and $N_f$ antiquark chiral superfields are grouped together and collectively denoted as $Q^a_i$, transforming in the fundamental representation of the enhanced flavor symmetry $SU(2N_f)$. Here $a=1,2$ is an $SU(2)$ color index, $i=1,\dots,2N_f$ is an $SU(2N_f)$ flavor index. The meson $M^{ij}=\epsilon^{ab}Q_a^i Q_b^j$, which comprises the usual mesons, baryons and antibaryons of $SU(N_c)$, transforms in the second rank antisymmetric representation $[0,1,0,\dots,0]$ of $SU(2N_f)$.

The charges of the quark $Q$, the meson $M$ and the monopole operator $Y\equiv V_{1}$ of minimal charge are listed in table \ref{tab:chargesSU2}.
\begin{table}[h]
\centering
\begin{tabular}{c| c c c}
         &  $U(1)_R$ & $U(1)_A$  &  $SU(2N_f)$ \\ \hline
$Q$    &   $r$  &  $1$   & $[1,0,0,\dots,0]$  \\ \hline
$M$    & $2r$  &  $2$   & $[0,1,0,\dots,0]$  \\
$\mathrm{Pf} M$  & $2N_f r$  &  $2N_f$   & $[0,\dots,0]$  \\
$Y$  & $2N_f(1-r)-2$  &  $-2N_f$   & $[0,\dots,0]$ \\ \hline
fugacities & $t$ & $y$  & $x$
\end{tabular}
\caption{Charges under the global symmetry for $SU(2)$ with $N_f$ flavors.}
\label{tab:chargesSU2}
\end{table}

The $R$-charge $r$ of the matter fields is taken to satisfy $0<r<1-\frac{1}{N_f}$, so that all gauge invariant chiral operators have positive $R$-charge. This requires $N_f>1$.

The nontrivial charges of monopole operators are 
\be\label{charges_monopoleSU2}
\begin{split}
R[V_{m}] &= \left( 2N_f(1-r)-2 \right) |m|  \\
A[V_{m}] &=-2N_f |m| ~,
\end{split}
\ee
where the magnetic charge $m$ denotes for the magnetic flux $\mathrm{diag}(m,-m)$ in $\mathfrak{su}(2)$.

The fundamental Weyl chamber of the GNO lattice for the magnetic charge $m$ is $\bZ/S_2=\bZ_{\geq 0}$. Since the instanton factor $Y^{-1}$ has $2N_f$ quark zero modes according to \eqref{quark_zero_modes}, there is no dynamically generated superpotential. Therefore the magnetic charge $m$ is summed over $\Gamma_q= \bZ_{\geq 0}$.

For $m=0$, the gauge group is unbroken and the contribution to the Hilbert series is the Hilbert series of the mesonic moduli space of $SU(2)$ with $N_f$ flavors, which is generated by the meson $M$ subject to the quadratic relation 
\be\label{relation_SU2_mes}
\epsilon_{i_1 i_2 i_3 i_4 \dots i_{2N_f}} M^{i_1 i_2} M^{i_3 i_4} = 0~,
\ee
that transforms in the representation $[0,0,0,1,0,\dots,0]$ of the flavor symmetry $SU(2N_f)$ and imposes $\mathrm{rk}(M)\leq 2$. Therefore, out of the $n$-th symmetric products of the antisymmetric meson $M$, only the representations $[0,n,0,\dots,0]$ survive in the chiral ring \cite{Gray:2008yu}. This mesonic Hilbert series is \cite{Gray:2008yu}
\be\label{SU2_mesonic}
H_{T_0}(t,y,x)= \sum_{n=0}^\infty [0,n,0,\dots,0]_x (t^r y)^{2n}~,
\ee
which reduces to 
\be\label{SU2_mesonic_ur}
H_{T_0}(t,y,1)= {}_{2}F_1(2N_f-1,2N_f;2;t^{2r} y^2)~
\ee
when unrefined with respect to the $SU(2N_f)$ fugacity $x$.

For nonvanishing magnetic charge $m>0$, the matter fields are massive and the residual gauge theory is trivial (a free $U(1)$ gauge theory). The sum simply counts bare monopole operators $V_{m}$, that contribute a weight $t^{(2N_f(1-r)-2)m} y^{-2N_f m}$. 

In total, the Hilbert series of the moduli space is therefore
\be\label{SU2_total}
H(t,y,x)= \sum_{n=0}^\infty [0,n,0,\dots,0]_x (t^r y)^{2n} + \frac{1}{1-t^{2N_f(1-r)-2} y^{-2N_f}} -1~.
\ee

The chiral ring is generated by the meson $M$ and the fundamental monopole operator $Y=V_{1}$, subject to the relations 
\be\label{relations_SU2}
Y M = 0~, \qquad \epsilon_{i_1 i_2 i_3 i_4 \dots  i_{2N_f}} M^{i_1 i_2} M^{i_3 i_4} = 0~.
\ee
The first relation says that the mesonic component and the Coulomb component, which is a copy of $\bC$ generated by $Y$, only meet at a point; the second relation is the relation \eqref{relation_SU2_mes} for the mesonic component. Even though it is not necessary, this can be explicitly confirmed by computing the plethystic logarithm of the Hilbert series \eqref{SU2_total} or by computing the Hilbert series of the ring generated by $Y$ and the meson $M$ subject to the relations \eqref{relations_SU2} using {\tt Macaulay2} to reproduce \eqref{SU2_total}.

For $N_f=2$, the chiral ring relations \eqref{relations_SU2} can be also obtained from a Wess-Zumino model in $Y$ and $M$ with superpotential $W=Y \pf M$ \cite{Aharony:1997bx}. Note that that even for $N_f>2$ the relations \eqref{relations_SU2} simply follow from the Coulomb-Higgs branch structure of the moduli space, which is implemented in the monopole formula for the Hilbert series. There was no need to use the singular effective superpotential $W_{eff} =(N_f-1)(Y \pf M)^{1/(N_f-1)}$ with the constraint \eqref{relation_SU2_mes} as in \cite{Aharony:1997bx}.

\section{$U(N)$ with $N_f\ge N$ flavors}\label{sec:U(N)_Nf}

Let us now consider $3d$ $\cN=2$ SQCD with gauge group $U(N)$ and $N_f\ge N$ flavors. This generalizes section \ref{sec:U(1)_flavors} to higher rank gauge groups. 
The charges of the quark $Q$, antiquark $\tilde{Q}$, meson $M$ and the fundamental monopole operators $V_+\equiv V_{(1,0,\dots,0)}$ and $V_-\equiv V_{(0,\dots,0,-1)}$ are listed in table \ref{tab:charges_U(N)_Nf}. 
\begin{table}[h]
\centering
\begin{tabular}{c| c c c c}
         &  $U(1)_R$ & $U(1)_A$  &  $SU(N_f)_L\times SU(N_f)_R$  & $U(1)_J$ \\ \hline
$Q$    &   $r$  &  $1$   & $[0,0,\dots,0,1;0,0,\dots,0,0]$ &   $0$  \\
$\tilde{Q}$  & $r$  &  $1$  & $[0,0,\dots,0,0;1,0,\dots,0,0]$  &   $0$  \\ \hline
$M$    & $2r$  &  $2$   & $[0,0,\dots,0,1;1,0,\dots,0,0]$ &   $0$  \\
$V_+$  & $N_f(1-r)-N+1$  &  $-N_f$ & $[0,0,\dots,0,0;0,0,\dots,0,0]$ &  $1$ \\
$V_-$  & $N_f(1-r)-N+1$  &  $-N_f$ & $[0,0,\dots,0,0;0,0,\dots,0,0]$ &  $-1$ \\ \hline
fugacities & $t$ & $y$  & $u~\qquad,\qquad~v$ & $z$
\end{tabular}
\caption{Charges under the global symmetry for $U(N)$ with $N_f$ flavors.}
\label{tab:charges_U(N)_Nf}
\end{table}

The $R$-charge $r$ of the matter fields is taken to satisfy $0<r<\frac{N_f+1-N}{N_f}$, which is possible for $N_f\geq N$, so that all gauge invariant chiral operators have positive $R$-charge. 

As reviewed in section \ref{subsec:lifting}, the $N$-dimensional classical Coulomb branch is lifted by non-perturbative corrections, except for a two-dimensional quantum Coulomb branch parametrized by $\sigma_1\geq \sigma_2=\dots = \sigma_{N-1}=0\geq \sigma_N$ \cite{Aharony:1997bx,deBoer:1997kr}. Correspondingly, the chiral monopole operators that parametrize the quantum Coulomb branch are labelled by magnetic charges $m$ in the two-dimensional quantum sublattice $\Gamma_q=\{m=(m_1,0,\dots,0,m_N) \in \bZ^N ~|~ m_1\geq 0 \geq m_N\}$. The non-trivial charges of these monopole operators 
\be\label{charges_monopole_UN}
\begin{split}
R[V_{(m_1,0,\dots,0,m_N)}] &= \left[N_f(1-r)-N+1\right](m_1-m_N) \\
A[V_{(m_1,0,\dots,0,m_N)}] &=-N_f(m_1-m_N) \\
J[V_{(m_1,0,\dots,0,m_N)}] &= m_1+m_N~
\end{split}
\ee
are compatible with the identification $V_{(m_1,0,\dots,0,m_N)}=V_+^{m_1} V_-^{-m_N}$. 

The residual gauge theory in the background of a monopole operator $V_{(m_1,0,\dots,0,m_N)}$ is non-trivial. There are four different cases, corresponding to the origin, the two one-dimensional boundary components and the interior of $\Gamma_q$:%
\footnote{Decoupled pure $U(1)$ gauge factors are omitted from the residual theories.}
\begin{enumerate}
\item $m_1=0=m_N$: the residual theory is the whole $U(N)$ with $N_f$ flavors; 
\item $m_1> 0=m_N$: the residual theory is $U(N-1)$ with $N_f$ flavors;
\item $m_1=0> m_N$: the residual theory is $U(N-1)$ with $N_f$ flavors; 
\item $m_1> 0 > m_N$: the residual theory is $U(N-2)$ with $N_f$ flavors. 
\end{enumerate}
The dressing factors due to these residual gauge theories 
\be\label{dressing_UN}
H_{T_m}(t,y,u,v)=H^U_{N_c(m),\,N_f}(t,y,u,v)
\ee
are mesonic Hilbert series $H^U_{N_c(m),\,N_f}(t,y,u,v)$ of $U(N_c)$ gauge theories with $N_f$ flavors, where the rank of the nontrivial part of the residual gauge group is $N_c(m)=N,N-1,N-2$ depending on the values of $m_1$, $m_{N_c}$ as explained above. The mesonic Hilbert series of $U(N_c)$ gauge theories with $N_f\geq N_c$ flavors can be obtained from those computed for $SU(N_c)$ gauge theories in \cite{Gray:2008yu} by projecting to operators with vanishing baryonic charge. They are given by
\be\label{mesonic_UNc}
\begin{split}
&H^U_{N_c,\,N_f}(t,y,u,v) = \\
&\quad =\sum_{n_1,\dots, n_{N_c}\geq 0} 
[0^{N_f-N_c-1},n_{N_c},\dots,n_1;n_1,\dots,n_{N_c},0^{N_f-N_c-1}]_{u,v} 
(t^r y)^{2\sum_{j=1}^{N_c} j n_j} ~,
\end{split}
\ee
where $0^n$ denotes a string of $n$ zeros.

Adding up these contributions weighted by the factors associated to bare monopole operators, the Hilbert series of the total moduli space is
\be\label{HS_U(N)_Nf_short}
\begin{split}
& H(t,y,u,v,z) = H^U_{N,\,N_f}(t,y,u,v) + \\
& \quad + \left[\frac{t^{N_f(1-r)-N+1}y^{-N_f} z}{1-t^{N_f(1-r)-N+1}y^{-N_f} z}+ \frac{t^{N_f(1-r)-N+1}y^{-N_f} z^{-1}}{1-t^{N_f(1-r)-N+1}y^{-N_f} z^{-1}}\right] H^U_{N-1,\,N_f}(t,y,u,v) + \\
& \quad + \left[\frac{t^{N_f(1-r)-N+1}y^{-N_f} z}{1-t^{N_f(1-r)-N+1}y^{-N_f} z} \frac{t^{N_f(1-r)-N+1}y^{-N_f} z^{-1}}{1-t^{N_f(1-r)-N+1}y^{-N_f} z^{-1}}\right] H^U_{N-2,\,N_f}(t,y,u,v)~.
\end{split}
\ee
The terms in brackets are produced by the sum over monopole operators, and those outside the brackets are the associated dressing factors of the form \eqref{mesonic_UNc}.

The discussion of the residual gauge theories and the counting of dressed monopole operators performed above shows that the ring of chiral gauge invariant operators which acquire VEV on the moduli space is generated by $V_+$, $V_-$ and the $N_f\times N_f$ meson matrix $M$, subject to the relations 
\be\label{relns_U(N)}
~\mathrm{minor}_{N+1}(M) = 0~,\quad V_{\pm} ~\mathrm{minor}_{N}(M) = 0~, \qquad V_+ V_- ~\mathrm{minor}_{N-1}(M) = 0~, 
\ee
where $\mathrm{minor}_{k}(M)$ denotes the set (or matrix) of $k\times k$ minors of $M$. The first relation describes the component 1 discussed above, and holds for the other components \emph{a fortiori}; the second set of relations describes components 2 and 3; the fourth relation describes component 4. 
Note that the dressed monopole operators can be expressed in terms of $V_+$, $V_-$ and $M$ using the relations \eqref{relns_U(N)}. There are no dressed monopole operators among the generators.

This result reproduces again the picture of the quantum moduli space obtained in \cite{Aharony:1997bx,Aharony:1997gp}, with no need of an effective superpotential.

\section{$USp(2N)$ with $N_f\ge N+1$ flavors}\label{sec:USp(2N)_Nf}

The analysis can be repeated for $3d$ $\cN=2$ SQCD with gauge group $USp(2N)$ and $N_f\ge N+1$ flavors, that is $2N_f$ fundamental chiral multiplets. For $N=1$ this reduces to the $SU(2)$ gauge theory with $N_f$ flavors of section \ref{sec:SU(2)_flavors}.
The charges of the quark $Q$, the meson $M$ and the fundamental monopole operator $Y \equiv V_{(1,0^{N-1})}$ are listed in table \ref{tab:charges_USp(2N)_Nf}. 
\begin{table}[h]
\centering
\begin{tabular}{c| c c c c}
         &  $U(1)_R$ & $U(1)_A$  &  $SU(2N_f)$  \\ \hline
$Q$    &   $r$  &  $1$   & $[1,0,0,\dots,0]$   \\ \hline
$M$    & $2r$  &  $2$   & $[0,1,0,\dots,0]$   \\
$Y$  & $2N_f(1-r)-2N$  &  $-2N_f$ & $[0,0,0,\dots,0]$   \\ \hline
fugacities & $t$ & $y$  & $x$ 
\end{tabular}
\caption{Charges under the global symmetry for $USp(2N)$ with $N_f$ flavors.}
\label{tab:charges_USp(2N)_Nf}
\end{table}

The $R$-charge $r$ of $Q$ is taken to satisfy $0<r<\frac{N_f-N}{N_f}$, which is possible for $N_f\geq N+1$, so that all gauge invariant chiral operators have positive $R$-charge. 

The analysis reviewed in section \ref{subsec:lifting} shows that the quantum Coulomb branch that survives non-perturbative corrections is one-dimensional and corresponds to $\sigma_1\geq \sigma_2=\dots = \sigma_{N}=0$ \cite{Karch:1997ux}. The chiral monopole operators that parametrize the quantum Coulomb branch are labelled by magnetic charges $m$ in the one-dimensional quantum sublattice $\Gamma_q=\{m=(m_1,0^{N-1}) \in \bZ^N ~|~ m_1\geq 0 \}$. The non-trivial charges of these monopole operators 
\be\label{charges_monopole_USp}
\begin{split}
R[V_{(m_1,0^{N-1})}] &= \left[2N_f(1-r)-2N\right]m_1 \\
A[V_{(m_1,0^{N-1})}] &=-2N_f m_1 
\end{split}
\ee
are compatible with the identification $V_{(m_1,0^{N-1})}=Y^{m_1}$. 

The residual gauge theory in the background of a monopole operator $V_{(m_1,0^{N-1})}$ is non-trivial. There are two different cases, corresponding to the origin (the boundary) and the interior of $\Gamma_q$:
\begin{enumerate}
\item $m_1=0$: the residual theory is the whole $USp(2N)$ with $N_f$ flavors; 
\item $m_1>0$: the residual theory is $USp(2(N-1))$ with $N_f$ flavors. 
\end{enumerate}
The dressing factors due to these residual gauge theories 
\be\label{dressing_USp}
H_{T_m}(t,y,x)=H^{USp}_{N_c(m),\,N_f}(t,y,x)
\ee
are mesonic Hilbert series $H^{USp}_{N_c(m),\,N_f}(t,y,x)$ of $USp(2N_c)$ gauge theories with $N_f$ flavors, where the rank of the nontrivial part of the residual gauge group is $N_c(m)=N,N-1$ depending on the value of $m_1$ as explained above. The mesonic Hilbert series of a $USp(2N_c)$ gauge theory with $N_f\geq N_c$ flavors is \cite{Hanany:2008kn}
\be\label{mesonic_USp}
\begin{split}
&H^{USp}_{N_c,\,N_f}(t,y,x) = \\
&\quad =\sum_{n_2,n_4,\dots, n_{2N_c}\geq 0} 
[0,n_2,0,n_4,\dots,0,n_{2N_c},0^{2(N_f-N_c)-1}]_{x} 
(t^r y)^{2\sum_{j=1}^{N_c} j n_{2j}} ~.
\end{split}
\ee
The associated chiral ring is generated by the mesons $M^{i j}$ subject to the relation 
\be\label{relation_USp_mes}
\epsilon_{i_1\dots i_{2N_f}} M^{i_1 i_2} \dots M^{i_{2N_c+1}i_{2N_c+2}}=0~,
\ee
that transforms in the representation $[0^{2N_c+1},1,0^{2(N_f-N_c)-3}]$ of the flavor symmetry and imposes $\mathrm{rk}(M)\leq 2N_c$, generalizing \eqref{relation_SU2_mes}.

Adding up these contributions to the Hilbert series, weighted by the factors associated to bare monopole operators, the Hilbert series of the moduli space is
\be\label{HS_USp(2N)_Nf_short}
H(t,y,x) = H^{USp}_{N,\,N_f}(t,y,x) +  \left[\frac{t^{2[N_f(1-r)-N]}y^{-2N_f}}{1-t^{2[N_f(1-r)-N]}y^{-2N_f}}\right] H^{USp}_{N-1,\,N_f}(t,y,x) ~.
\ee
The term in brackets is produced by the sum over monopole operators with $m_1>0$, and those outside the brackets are the mesonic dressing factors of the form \eqref{mesonic_USp}.

This analysis shows that the chiral ring is generated by the antisymmetric $2N$ by $2N$ meson matrix $M$ and by the fundamental monopole operator $Y$, subject to the rank relations 
\be\label{relns_USp}
\epsilon_{i_1\dots i_{2N_f}} M^{i_1 i_2} \dots M^{i_{2N+1}i_{2N+2}}=0~, \quad Y \cdot \epsilon_{i_1\dots i_{2N_f}} M^{i_1 i_2} \dots M^{i_{2N-1}i_{2N}}=0~.
\ee
This result reproduces the picture of the quantum moduli space obtained in \cite{Karch:1997ux}.

\section{Conclusions} \label{sec:_conclusions}

In this work I presented a simple monopole formula \eqref{monopole_formula} for the Hilbert series that counts gauge invariant chiral operators in 3d $\cN=2$ Yang-Mills theories with vectorlike matter. The formula is a generalization of the monopole formula for the Coulomb branch of 3d $\cN=4$ gauge theories introduced in \cite{Cremonesi:2013lqa} and counts dressed 't Hooft monopole operators. In $\cN=2$ theories, the dressing is due to gauge invariant operators of the residual gauge theory of massless vector and chiral multiplets in the monopole background (or equivalently on the Coulomb branch). The sum over magnetic charges is restricted to a sublattice of the fundamental Weyl chamber of the GNO lattice, to take into account the partial lifting of the classical Coulomb branch by instanton generated superpotentials.

I computed the Hilbert series for unitary and symplectic gauge theories with fundamental matter, reproducing old results on their moduli spaces of supersymmetric vacua and chiral rings. Unlike the old semiclassical analysis, the new formalism does not rely on any effective superpotential that depends on the generators of the chiral ring, such as monopole operators and mesons. The quantum relations involving monopole operators are an outcome of the formalism and simply follow from the Coulomb-Higgs fibration structure of the total moduli space. 

It would be interesting to apply this formalism to $\cN=2$ Yang-Mills theories with other gauge groups and matter content, without or with superpotential, and use it to test dualities. The computation could be particularly instructive for theories which are expected to have a richer chiral ring including dressed monopole operators among the generators, such as adjoint SQCD theories \cite{Kim:2013cma}. 

The picture advocated in this paper also applies to the total moduli space of 3d $\cN=4$ good or ugly gauge theories, which is the union of a Higgs branch (see \cite{Hanany:2011db} for the Hilbert series), a Coulomb branch, and mixed branches. Due to $\cN=4$ supersymmetry there are no instanton corrections to the superpotential, therefore the magnetic charges span the whole fundamental Weyl chamber in the GNO lattice. The dressing of bare monopole operators involves chiral operators arising from both $\cN=4$ vector multiplets and $\cN=4$ hypermultiplets, when decomposed in $\cN=2$ multiplets, further subject to $F$-term equations implied by the superpotential. It would be interesting to compute the Hilbert series of the total vacuum moduli space of $T[SU(N)]$ theories \cite{Gaiotto:2008ak} and compare the result with the structure of the moduli space conjectured in \cite{Chacaltana:2012zy}.

\section*{Acknowledgements} I thank Amihay Hanany for suggesting the problem, and Noppadol Mekareeya and Alberto Zaffaroni for collaboration on a related project and comments on a preliminary version of the draft. I am particularly grateful to Alberto Zaffaroni for several insightful discussions.

\bibliographystyle{utphys}
\bibliography{references}{}

\end{document}